\documentclass[%
preprint,
 amsmath,amssymb,
 aip,
jap,
floatfix,
]{revtex4-1}

\usepackage{graphicx}
\usepackage{dcolumn}
\usepackage{bm}
\begin{document}

\preprint{AIP/JAP}

\title{Tune of Magnetism and Electronic Structures of Alkali Metal Carbides with Rocksalt Structure}
\author{Wenxu Zhang}\email{xwzhang@uestc.edu.cn}
\author{Zhida Song, Bin Peng, Wanli Zhang}
\affiliation{State Key Laboratory of Electronic Thin Films and Integrated Devices, University of
Electronic Science and Technology of China, Chengdu 610054, China}

\date{\today}

\begin{abstract}
Electronic structures of carbides with the rocksalt structure were calculated by full potential electronic codes solving the Kohn-Sham equation. Bonding characters were analyzed by constructing tight-binding Hamiltonian based on maximally-localized Wannier functions. It was found that the cations in these compounds act as an electron provider and the frame is formed by the carbon atoms. The electronic states in the vicinity of the Fermi level are mainly from the p-orbitals of C. Pressure and doping are two efficient ways to tune the magnetic and electronic properties of these compounds. It turns out that a spin gapless semiconductor can be obtained by applying hydrostatic pressure up to tens of gigaPascal. Higher pressure induced an insulator to metal transition because of band broadening. Compounds of IA group (Na, K, Rb, Cs) were magnetic semiconductor at ambient conditions. Alloying with IIA elements decrease the magnetic moment according to the law of $3-x$, where $x$ is the relative atomic ratio of the IIA elements to the IA ones. The behaviors of the compounds under the pressure and the doping effects can be understood by a rigid band model.
\end{abstract}

\maketitle
\section{Introduction}
Magnetic materials are essential for modern technology. Up to date, most used magnetic materials involve elements with partially filled d or f-orbitals, like transitional metals, rare earth metals and their compounds. Magnetic ordering in these compounds originates from the localization of the electrons so that the kinetic energy can be minimized. This is the result of competitions between the kinetic energy and the exchange splitting. The former favors nonmagnetic state while the latter is the source of spin polarization.
\par Magnetism from p-states rather than the above conventional d or f states has been a hot topic in recent years because of technical importance and theoretical interests as well. There are several ways to introduce spin polarization to the p-states. Several alkali and alkaline earth compounds with elements from group IVA (carbides and silicides),\cite{verma} VA(nitrides, pnictides, and arsenides)\cite{gao09,ozdogan,sieberer,Yan}, and VIA (selenlides)\cite{gao11,Yogeswari} are predicted to be magnetically ordered. Oxides are not so usual, but doping with C and N, or introducing cation vacancies can be sources of magnetism.\cite{pardo} Another large category comes from nanostructures of graphene, graphite and C$_{60}$\cite{yazyev} or their modification such as doping or absorption of other atoms. The origin of the spin polarization is explained mostly from the Stoner's theory where localization of the p-orbitals triggers the instability of the nonmagnetic state to magnetic state\cite{volnianska}. Bulk carbon compounds are rarely found to be spin polarized.
 \par The most distinct feature of these compounds with p-orbital magnetism is that in most cases they are half metals, where in the one electron picture, one spin channel in the Kohn-Sham band structure has a finite density of states(DOS), while the other is null. This is very important and promising for its possible applications in spintronics.\cite{strmp}
\par Monocarbides MC, where M may be alkali metal and alkaline earth metal show the above interesting magnetic properties, namely half metallic ferromagnetism, due to there unique electronic structures as predicted by several theoretical works.\cite{Gao07,Gao07prb,zhang08,dong11,zhang12} In our previous work,\cite{zhang12} ferromagnetic ordering and its electronic origin were investigated in alkaline earth carbides with the rocksalt structure. It was found that the magnetism is from the polarization of the carbon p-orbitals, while the alkaline earth metals, due to their large atomic radii, guarantee the sufficient large distance between the carbons, which localizes the p-orbitals. At the same time, they are also electron provider because of the smaller electronegativity than that of C. In this case, we proposed that the electronic structure and the spin magnetic moment can be nicely tuned, which gives us freedom to tailor their properties when used as a spin filter, or spin injector. In this work, we show how the electronic structure can be tailored by doping (A$^{IA}_x$B$^{IIA}_{1-x}$C) and by hydrostatic pressures. It turns out that the fundamental gap as well as the spin flip gap can be tuned and induce the transition from a semiconductor to metal.

\section{Calculation details}
The calculations were performed by the
full-potential local-orbital code\cite{koep} in the version
FPLO9.00-33 with the default basis settings. All calculations
were done within the scalar relativistic approximation. The general gradient approximation of exchange-correlation
functional was chosen to be that parameterized by Perdew, Burke, and Ernzerhof.\cite{gga}
The number of k-points in the whole Brillouin zone was set to $32\times32\times32$
in order to ensure the convergency of the results. Convergency of the total energy was set to be better than $10^{-8}$ Hartree together with that of the electron density better than $10^{-6}$ in the internal unit of the codes. The FCC unit cell (space group Fm$\overline{3}$m) was used to represent the rocksalt structure with Wyckoff positions assigned to metal (M) and C as M(0,0,0) and C(0.5,0.5,0.5), respectively. Doping was modeled in a supercell with the total atom number up to 32.

\section{Electronic structure of the compounds}

\begin{figure}
\includegraphics[scale=0.45]{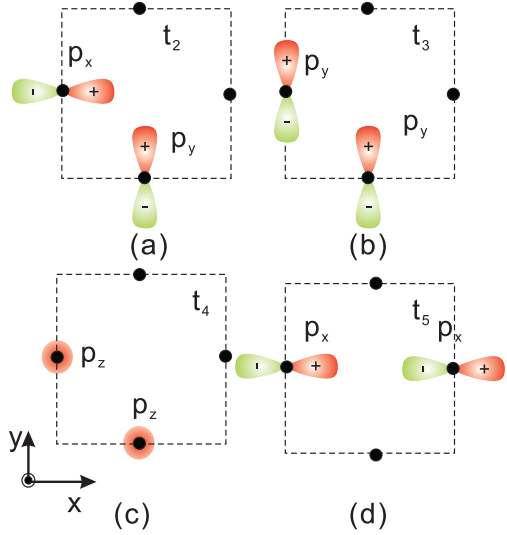}
\caption{\label{fig:tpp} (Color online) Cartoon representation of the interactions between different p-like WFs centered on the carbons.}
\end{figure}

\begin{figure}
\includegraphics[scale=0.45]{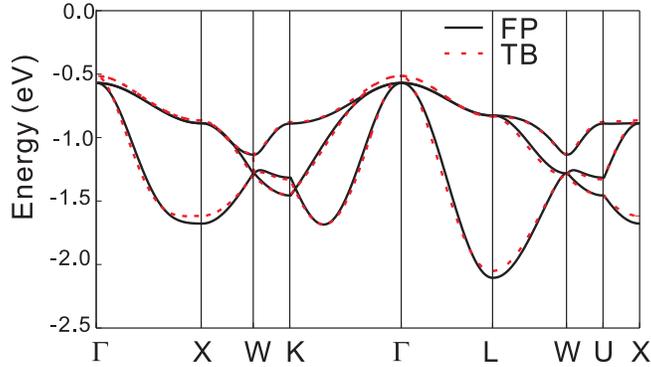}
\caption{\label{fig:wfbands} (Color online) Comparison of the bands from TB model (TB dashed curves) and full potential calculations(FP solid curves).}
\end{figure}

\begin{figure}
\includegraphics[scale=0.45]{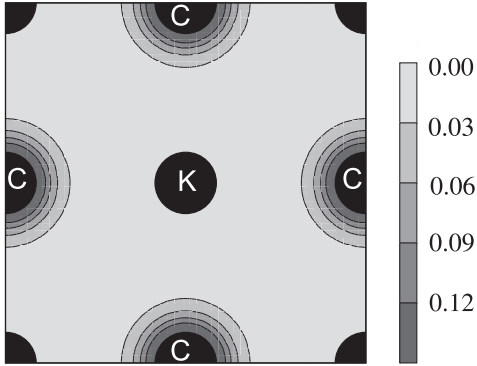}
\caption{\label{fig:edensity} The contour lines of the electron charge density of KC at equilibrium lattice constant.}
\end{figure}

As found in our previous work,\cite{zhang12} the DOS around the Fermi level are mainly contributed by the 2p-electrons of C. We constructed 3 carbon centered Wannier functions(WFs). The resulting WFs are p-like.
In the basis of these WFs, if the small interactions below 0.001 eV are ignored, the elements of the tight-binding(TB) non-magnetic Hamiltonian $[H_{ij}]_{3\times3}$ can be written as,
\begin{equation}\label{hii}
    H_{ii}=t_1+4t_3\cos(\frac{a}{2}k_i)[\cos(\frac{a}{2}k_j)+\cos(\frac{a}{2}k_k)]+4t_4\cos(\frac{a}{2}k_j)\cos(\frac{a}{2}k_k)+2t_5\cos(ak_i)
\end{equation}
 and
 \begin{equation}
 H_{ij}=H_{ji}=-4t_2\sin(\frac{a}{2}k_i)\sin(\frac{a}{2}k_j)
 \end{equation}
 where $a$ is the lattice constant, and subscripts $i,j$and$k=x,y$ and $z$, respectively. The hoping constants $t_1$ is onsite contribution,  $t_2$, $t_3$ and $t_4$ are from the nearest neighbor interactions, and $t_5$ is from the next-nearest neighbor interactions as shown in Fig. \ref{fig:tpp}. The values are $t_1=-1.13$ eV, $t_2=0.11$ eV, $t_3=0.07$ eV, $t_4=-0.03$ eV and $t_5=0.06$ eV. We see that the strongest interactions come from neighboring carbons with orthogonal p-orbitals. The bands from this TB Hamiltonian are shown in Fig. \ref{fig:wfbands} together with those from the full potential calculations. The agreement of the two results is quite satisfactory.
In order to illustrate the binding in the real space, the charge density contour of the compounds is shown in Fig. \ref{fig:edensity}. Spherical distribution of the electrons around the ions is a typical case of ionic bonding between the metals and carbon. In this scenario, the metallic ions determines the distances between the carbons. The characteristics of electrons around the Fermi level are mostly determined by the interaction of the 2p-orbitals from C. Exchange interaction is thus direct exchange between the carbons like in zinc-blende M$^{IIA}$C\cite{Gao07prb}.
\par This specific electronic structure gives us a large degree of freedom to tune the properties of the material as will be discussed in the following sections.

\section{Pressure dependence of the electronic structure and magnetism}
\begin{figure}
\includegraphics[scale=0.45]{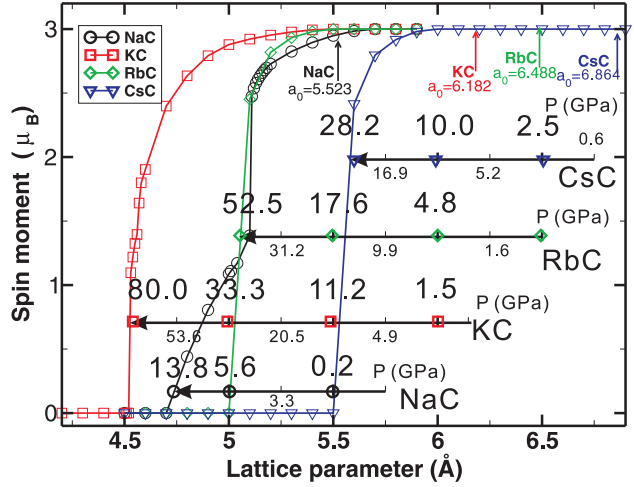}
\caption{\label{fig:SSIA} Dependence of the spin magnetic moment per primitive
cell on the lattice constant. The corresponding pressure at the different lattice constant is also shown. The equilibrium lattice constant $a_0$ of the compounds are marked with arrows.}
\end{figure}

\begin{figure}
\includegraphics[scale=0.45]{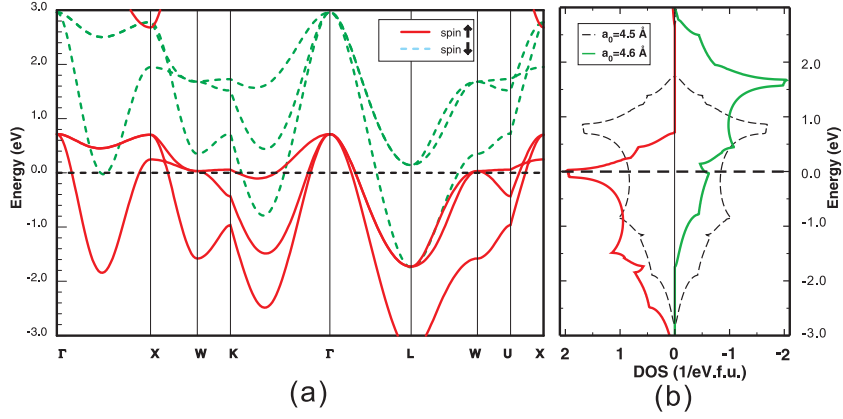}
\caption{\label{fig:dosKC}(Color online) Band of KC at lattice constant $a=4.6$ \AA\ (a), and DOS of KC at lattice constant $a=4.5$ \AA\ and 4.6 \AA\ (b). The minus sign in the DOS indicates that it is of spin down channel, while the numbers without any signs indicate that they are of the spin up channel. The following DOS figures follow this convention if not specified.}
\end{figure}

\begin{figure}
\includegraphics[scale=0.5]{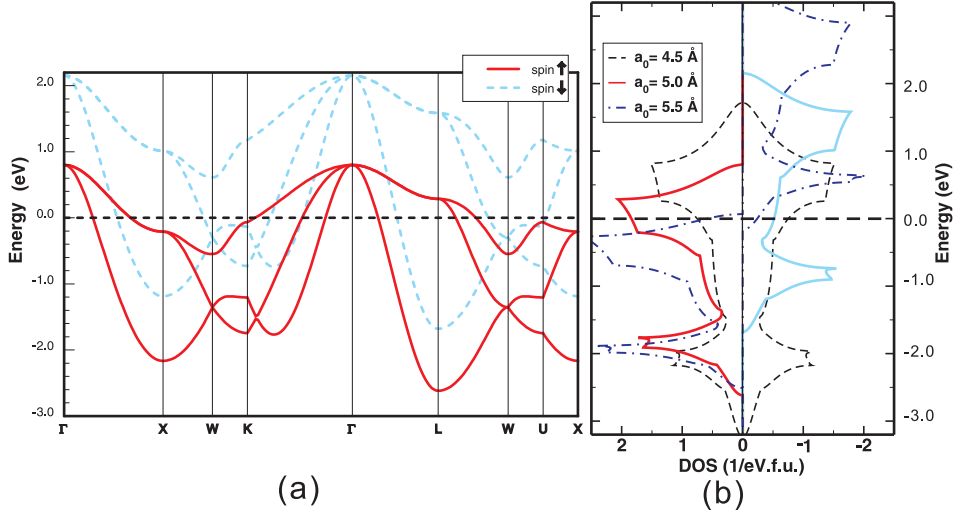}
\caption{\label{fig:dosNaC} (Color online) Band of NaC at lattice constant $a=5.0$ \AA\ (a), and DOS of NaC at lattice constant $a=4.5$ \AA, 5.0\AA\ and 5.5 \AA\ (b). }
\end{figure}

As changes of the volume can result in changes of the bandwidth, the DOS of these
compounds are strongly dependent on the pressure which results in the dependence
of magnetism on the volume (pressure).
The dependence of the spin magnetic moment on the volume of M$^{IA}$C is illustrated in Fig. \ref{fig:SSIA}.
It can be clearly seen that shrinking of the lattice constant can decrease the spin magnetic moment in all the compounds. This is the results of the increased kinetic energy of the electrons under pressure as expected. The lattice constant at which the spin polarization is set in is proportional to that at equilibrium, except that of NaC which is close to the lattice constant at equilibrium. This is because NaC is metallic at equilibrium as shown in our previous work \cite{zhang12}. However, the processes of the magnetic moment approaching zero are different: The spin moment of KC, RbC and CsC show an abrupt drop to zero at certain critical lattice constant while that of NaC drops first to an intermediate value of 1.5 $\mu_B$ and then to zero. The difference comes from the different DOS at the Fermi level. By simple consideration from the rigid band model, the susceptibility of the magnetic moment to external parameters is proportional to the DOS at the Fermi level and its derivative. The DOS of NaC and KC in the vicinity of the transitional region are shown in Fig. \ref{fig:dosKC} and Fig. \ref{fig:dosNaC} in order to illustrate this point. Comparing the DOS of KC at lattice constant of 4.5 and 4.6 \AA\ in Fig. \ref{fig:dosKC}, we can see that a rigid shift of the states of spin up and spin down is obvious in the spin polarized state. In KC, the onset of spin polarization at 4.6 \AA\ set the Fermi level at the high DOS about 2 eV$^{-1}$ in the spin up channel. The Fermi level is  between two saddle points. In NaC, the Fermi level is also between the two saddle points when the compound is at the intermediate magnetic moment state (e.g. 1.2 $\mu_B$ when $a=5.0$ \AA) as shown in Fig. \ref{fig:dosNaC}. The spin polarization at this lattice constant is because of the high DOS at the Fermi level in the nonmagnetic state (not shown) which triggers the magnetic instability according to the Stoner's model. Increase of the lattice constant enhances the exchange splitting, and the Fermi level passes through the upper saddle point as shown in Fig. \ref{fig:dosNaC}(b). In this case the magnetic moment increase quickly to a higher value.

\par The critical pressure of phase transition from the ferromagnetic (FM) state to the nonmagnetic (NM) state of
CsC and NaC is approximately 28.2 GPa and 13.8 GPa. The FM state of KC is so stable and begins to decrease at about 30 GPa. Under high pressure over 80 GPa we see the transition of the compound transforms into the NM state.

\par
\begin{figure}
\includegraphics[scale=0.45]{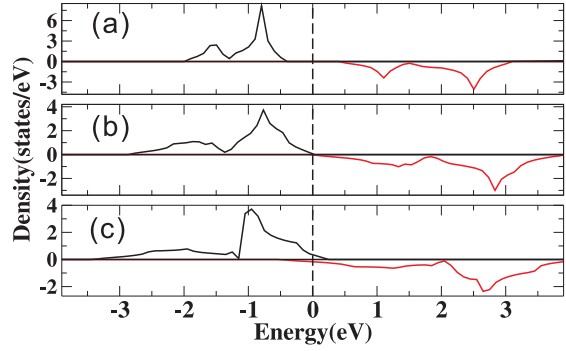}
\caption{\label{fig:DOS_KC} DOS of KC at different lattice constant: (a) 6.21, (b) 5.18 and (c) 4.76 \AA\, respectively.}
\end{figure}

 Not only the magnetic order, but also the band gap can be tuned by pressure. When the compounds are semiconductor with a finite gap, the bandwidth starts to broaden under pressure. In this case the bandgap decreases continuously. Normally insulating states transforms into metallic states.  An example from KC is shown in Fig. \ref{fig:DOS_KC}. At zero pressure, corresponding to the lattice constant $a=6.21$\AA\, the gap is 1.0 eV formed between the two spin states as in Fig. \ref{fig:DOS_KC}(a). It is called the spin flip gap as noted by Capelle \cite{capelle}. The fundamental gaps are the same as the spin flip gap. From the band structure, we also find that the gap is an indirect gap between the $\Gamma$ and $L-$point in the Brillouin zone. Under pressure, the gap comes to closure as shown in Fig. \ref{fig:DOS_KC}(c). The compound transforms from magnetic insulator to magnetic metal.
At the some critical pressure, the highest occupied spin up states just touches the lowest unoccupied spin down states as in Fig.\ref{fig:DOS_KC}(b). It reaches an interesting situation where no energy is required to flip the spin. This spin-gapless semiconductor has some important features including that the electron
excitation from valence bands to conduction bands needs no energy and the carriers
are fully spin-polarized. As the carriers are fully spin-polarized, they can be easily separated by Hall
effect. The features are useful in some novel applications as recently reviewed by Wang\cite{wang} \emph{et al}.

\begin{figure}
\includegraphics[scale=0.45]{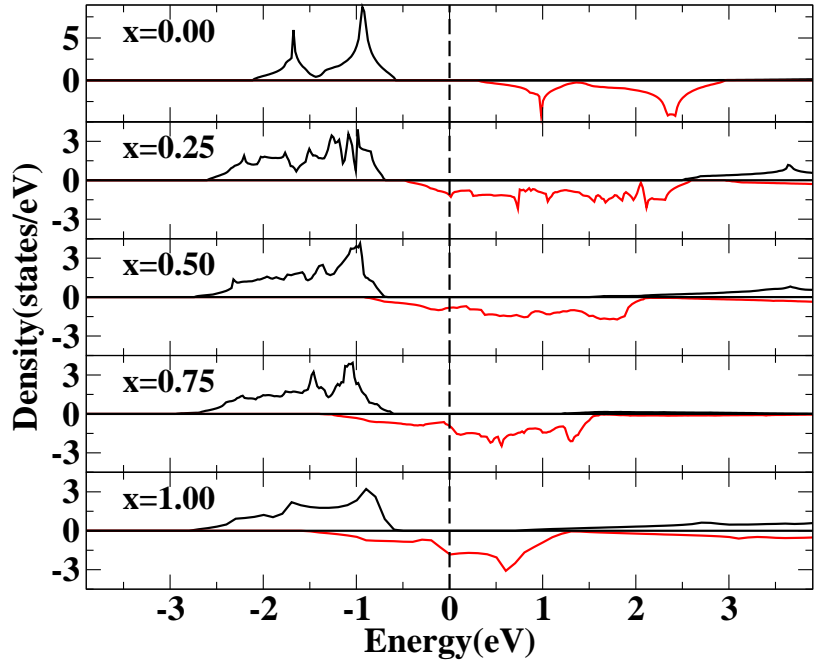}
\caption{\label{fig:DOS_KBaC} DOS of K$_x$Ba$_{1-x}$C at different doping levels indicated by $x$.}
\end{figure}

\section{Tune of the magnetic moment by atomic substitution}
As was shown above, the electronic states near the Fermi level are from the p-orbitals of carbon. The alkali metals serve as electron provider which determines the Fermi level of the system. We already obtained that BaC and KC have magnetic moment of 2.0 and 3.0 $\mu_B$, respectively. In these two compounds, the spin up channel are fully occupied. In this case, the change of the electron number will only change the occupation of the spin down channel. If we substitute the IA elements by IIA ones, forming compounds M$^{IA}_x$K$^{IIA}_{1-x}$C, the magnetic moment $M$ varies as $M=3-x$.
\par The variation of the DOS and the magnetic moment at different doping levels when K is substituted by Ba in KC are shown in Fig. \ref{fig:DOS_KBaC}. Starting from KC, the spin up states are fully occupied while the spin down states are empty, because there are 3 valence electrons and 3 p-states from C are available. Doped with the IIA element Ba, for instance, the extra electrons can only go to the spin down channel. Then the metallic state is produced and the electrons at the the Fermi level is fully spin polarized. Thus, a half metal is formed. The magnetic moment decreases according to $M=3-x$. At the same time, the half metallicity preserves. Thus a tunable half metal is obtained. The decrease of the magnetic moment can benefit devices where magnetic static interferences can be reduced.
\section{Conclusion}
In summary, we performed electronic structure calculations of the MC compounds where $M$ is one of the IA elements (Na, K, Rb, and Cs) based on DFT. The results show that there is direct exchange of p-electrons of C which is the source of magnetism. The alkali metals serve as an electron source. The states near the Fermi level are contributed mainly from carbon. All the compounds are ferromagnetic insulator at equilibrium with spin magnetic moment 3.0 $\mu_B$, except that NaC is metallic. NaC can be insulator by slight expansion of the lattice constant. The electronic structure and magnetism of the compounds can be tuned by pressure and doping with alkaline earth metals. The variation of the physical parameters can be understood by the rigid band model. The compounds can have potential applications in spintronics and photoelectronics dure to its half metallicity and spin gapless feature.
\section{Acknowledgement}
This work was financially supported by International Science \& Technology Cooperation Program of China (2012DFA51430) and the Fundamental Research Funds for the Central Universities (ZYGX2011J021).

\end{document}